\documentclass{emulateapj}

\newcommand{\eps}[1]{\mbox{log~$\epsilon$\,(#1)}}
\newcommand\iso[2]{$^{\rm #1}$#2}

\def\bd{\mbox{BD~$+$17~3248}}
\def\hd{\mbox{HD~122563}}

\def\kmsec{\mbox{km~s$^{\rm -1}$}}

\def\rpro{\mbox{$r$-process}}
\def\spro{\mbox{$s$-process}}
\def\ncap{\mbox{$n$-capture}}

\shorttitle{UV $n$-capture Abundances in BD~$+$17~3248}
\shortauthors{Roederer et al.}
\submitted{Accepted for publication in the Astrophysical Journal (Part 2)}

\begin{document}

\title{
New Abundance Determinations of Cadmium, Lutetium, and Osmium \\ in the 
$r$-process Enriched Star BD~$+$17~3248}

\author{
Ian U.\ Roederer\altaffilmark{1},
Christopher Sneden\altaffilmark{1},
James E.\ Lawler\altaffilmark{2}, and
John J.\ Cowan\altaffilmark{3}
}

\altaffiltext{1}{Department of Astronomy, University of Texas at Austin,
1 University Station, C1400, Austin, TX 78712-0259, USA; 
iur@astro.as.utexas.edu}

\altaffiltext{2}{Department of Physics, University of Wisconsin, 
Madison, WI 53706, USA}

\altaffiltext{3}{Homer L.\ Dodge Department of Physics and Astronomy, 
University of Oklahoma, Norman, OK 73019, USA}

\begin{abstract}

We report the detection
of Cd~\textsc{i} ($Z =$~48), 
Lu~\textsc{ii} ($Z =$~71), and Os~\textsc{ii} ($Z =$~76)
in the metal-poor star BD~$+$17~3248.
These abundances are derived from an 
ultraviolet spectrum obtained with the 
Space Telescope Imaging Spectrograph on the 
\textit{Hubble Space Telescope}.
This is the first detection of these neutron-capture
species in a metal-poor star enriched by the $r$-process.
We supplement these measurements with new abundances of
Mo~\textsc{i}, Ru~\textsc{i}, and Rh~\textsc{i} derived
from an optical spectrum obtained with the 
High Resolution Echelle Spectrograph on Keck.
Combined with previous abundance derivations,
32 neutron-capture elements have been detected in BD~$+$17~3248,
the most complete neutron-capture abundance pattern in any 
metal-poor star to date.
The light neutron-capture elements (38~$\leq Z \leq$~48)
show a more pronounced even-odd effect than expected from
current Solar system $r$-process abundance predictions.
The age for \bd\ derived from the Th~\textsc{ii}/Os~\textsc{ii}
chronometer is in better agreement with the age derived
from other chronometers than the age derived from
Th~\textsc{ii}/Os~\textsc{i}.
New Hf~\textsc{ii} abundance derivations from transitions
in the ultraviolet are lower than those derived from transitions
in the optical, and the lower Hf abundance is in 
better agreement with the scaled Solar system $r$-process distribution.

\end{abstract}


\section{Introduction}
\label{intro}

Steady progress has been made over the last half-century
toward understanding how the heaviest elements
in the Universe are produced.
For the elements heavier than the iron (Fe) group, the vast majority
of isotopes are produced by the successive addition of neutrons
to existing nuclei on timescales that are slow or rapid
relative to the average $\beta^{-}$ decay rates.
These are referred to as the slow ($s$) and rapid ($r$) neutron ($n$)
capture processes, respectively
(see, e.g., \citealt{truran02} and \citealt{sneden08} for 
discussion of these processes).
The basic physical principles of these reactions are well known.
The \spro\ involves isotopes near the valley of
$\beta$ stability, so the properties relevant to understanding 
the nature of the \spro\ 
(e.g., \ncap\ cross sections, half-lives, etc.) can be studied 
in laboratories on Earth (see \citealt{cowan91} and references therein).
Phenomelogical or 
nuclear reaction models can then be constructed
to predict the general abundance pattern produced by the \spro\
(see \citealt{busso99}).
When applied to the Solar system (S.~S.) heavy element abundance distribution,
the \spro\ abundances can be subtracted
from the total abundances to reveal the \rpro\ 
component (e.g., \citealt{seeger65,kappeler89,arlandini99}).
Due to the more energetic nature of the \rpro\ and the exotic, 
short-lived nuclei involved, reaction networks for the
\rpro\ were not tractable until only recently
(see \citealt{kratz07}).
To evaluate and verify detailed nucleosynthesis models,
abundance patterns must be accurately characterized for 
as many elements as possible in locations beyond the S.~S.

In this Letter, we report abundance estimates for 
neutral cadmium (Cd~\textsc{i}, $Z =$~48), 
singly ionized lutetium (Lu~\textsc{ii}, $Z =$~71), 
and singly ionized osmium (Os~\textsc{ii}, $Z =$~76)
in the near-ultraviolet (NUV) 
spectrum of the \rpro\ enriched metal-poor star \bd.
This is the first clear detection of Cd and Lu in 
a metal-poor star enriched by the \rpro.
Combined with previous abundance derivations \citep{cowan02,cowan05,sneden09}
and several other new abundances derived from the optical spectrum
of this star,
32 \ncap\ elements have been detected in \bd,
the most complete \ncap\ pattern in any metal-poor star.
In the metal-poor star \hd, which
is relatively deficient in the heavy \ncap\ elements,
we also report tentative detections for Cd~\textsc{i} 
and Lu~\textsc{ii}, as well as an upper limit for Os~\textsc{ii}.
Finally, we use these new abundances to differentiate among the
various techniques used to predict the \rpro\ abundance pattern.

\section{Observations and Abundance Analysis}
\label{analysis}

NUV spectra of \bd\ and \hd\
were obtained using the Space Telescope Imaging Spectrograph (STIS)
on the \textit{Hubble Space Telescope} (\textit{HST}). 
These spectra cover a wavelength region from 2280--3120\AA\ at 
$R \equiv \lambda/\Delta\lambda \sim$~30,000.
The optical spectrum of \bd\ was obtained using 
the High Resolution Echelle Spectrograph (HIRES; \citealt{vogt94})
on Keck~I, and this spectrum covers a 
wavelength region from 3120--4640\AA\ at
$R \sim$~45,000.
See \citet{cowan05} for further details.

\begin{figure}
\begin{center}
\includegraphics[angle=0,width=3.4in]{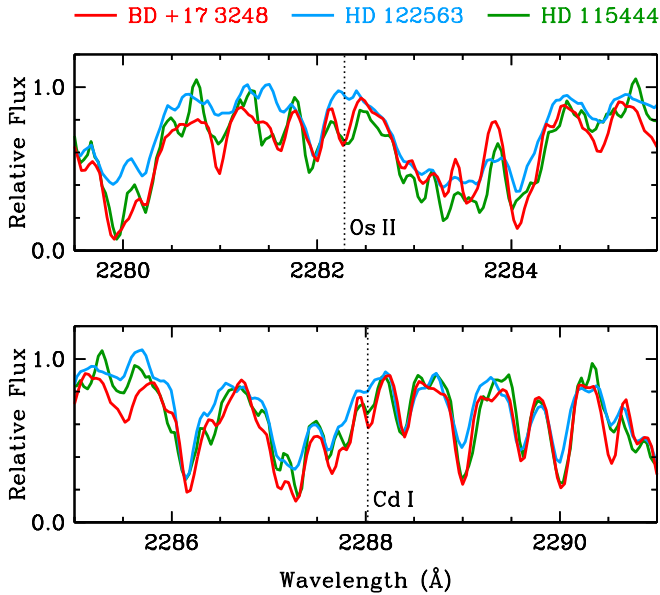}
\end{center}
\caption{
\label{overplot}
Spectral regions of \bd, \hd, and \mbox{HD~115444} 
surrounding the Os~\textsc{ii} and Cd~\textsc{i} lines.
The spectra have been smoothed to increase their signal-to-noise 
(S/N) ratios.
}
\end{figure}

In Figure~\ref{overplot}, we show segments of the STIS spectra
surrounding the Os~\textsc{ii} transition at 2282.28\AA\ and the
the Cd~\textsc{i} transition at 2288.02\AA\
in \bd\ and \hd, as well as \mbox{HD~115444}.\footnote{
This spectrum of \mbox{HD~115444}, taken with STIS using the 
same setup as the spectra of \bd\ and \hd,
has considerably lower S/N, and we 
do not examine \mbox{HD~115444} beyond this initial test.}
A strong absorption feature is clearly identified at these
wavelengths in \bd\ but not in \hd.
\bd\ is warmer ($T_{\rm eff} =$~5200~K)
and more metal-rich ([Fe/H]~$= -$2.1) than \hd\ 
($T_{\rm eff} =$~4570~K and [Fe/H]~$= -$2.7).
\mbox{HD~115444} has a temperature ($T_{\rm eff} =$~4720~K),
metallicity ([Fe/H]~$= -$2.9), and overall light element
abundance distribution (i.e., 6~$\leq Z \leq$~40)
that closely resembles \hd\ \citep{westin00}.
\mbox{HD~115444} is overabundant in the heavy
\ncap\ elements ([Eu/Fe]~$= +$0.7) relative to \hd\
([Eu/Fe]~$= -$0.5).
Therefore the only significant difference between the spectra of
\mbox{HD~115444} and \hd\ should be the stronger heavy \ncap\
absorption lines in \mbox{HD~115444}.
In Figure~\ref{overplot}, we see that \mbox{HD~115444}, like \bd, 
also exhibits strong absorption features at 2282.28 and 2288.02\AA,
but \hd\ does not.
Thus heavy \ncap\ species must be producing this absorption.
We find no transitions of heavy \ncap\ species 
at these wavelengths---or the Lu~\textsc{ii} line
at 2615.41\AA---in the Kurucz or NIST line databases that could plausibly
account for this absorption other than the species of interest.

References for published transition probabilities of the lines
used in this analysis are given in Table~\ref{abundtab}.
We determined the transition probability of the 
Lu~\textsc{ii} 2615.42\AA\
resonance line to be log($gf$)~$ = +$0.11~$\pm$~0.04 based on a
laser-induced fluorescence lifetime measurement of its upper level
\citep{fedchak00} and a branching fraction calculation of 0.971
\citep{quinet99}.  
(See also \citealt{lawler09}.)
The \iso{175}{Lu} isotope is dominant (97.4\% of S.~S.\ Lu;
\citealt{lodders03}). 
The \iso{176}{Lu} isotope is blocked from \rpro\ 
production by the stable \iso{176}{Yb} isotope, so it is
expected to be entirely absent from \bd. 
The odd-$Z$ isotope \iso{175}{Lu} has nonzero nuclear spin $I =$~7/2.
Hyperfine structure (hfs) and an accurate line position are based on
new laboratory measurements of the 6s6p $^{1}$P$^{0}$ 
level energy, 38223.406(8)~cm$^{-1}$, 
hfs A, $-$0.03731(10)~cm$^{-1}$, and 
hfs B, 0.0811(15)~cm$^{-1}$.
The naturally occurring \rpro\ isotopes of Cd and Os are predominantly
even-$Z$ even-$N$ isotopes with zero nuclear spin,
thus we are justified in ignoring the hfs from their
minority isotopes.

We use the current version of the LTE spectral analysis code MOOG
\citep{sneden73} to perform the abundance analysis.
We adopt the atmospheric parameters for \bd\ and \hd\ derived by
\citet{cowan02} and \citet{simmerer04}
($T_{\rm eff}$/log~$g$/[M/H]/$v_{t} =$ 
5200~K/1.80/$-$2.08/1.9~\kmsec\ and 
4570~K/1.35/$-$2.50/2.9~\kmsec, respectively)
and interpolate model atmospheres from the Kurucz grids
\citep{castelli97}.

We compare our results to abundances of other species 
derived from lines in the optical spectral range.
In the NUV, bound-free continuous opacity from metals may be comparable
to or greater than the bound-free continuous opacity from H$^{-}$
that dominates in the optical spectral range for metal-poor stars
(e.g., \citealt{travis68}).
To compensate for deficiencies in our ability to model the continuous
opacity in this spectral range, 
we have derived abundances of relatively clean, unsaturated, and 
unblended Fe~\textsc{i} and
Zr~\textsc{ii} lines across the NUV.
We require that these lines have reliable log($gf$) values
(Fe~\textsc{i}: \citealt{obrian91} or a grade of ``C'' or better in 
the NIST database; 
Zr~\textsc{ii}: \citealt{malcheva06}), and we derive the abundances
by matching synthetic to observed spectra.
Ideally we should select metals that make significant contributions
to the continuous opacity (e.g., Mg, as advocated by \citealt{bell01}), 
but practically we are constrained 
because there are very few metals whose lines
have reliable laboratory transition probabilities
and are unsaturated and unblended in the NUV spectra
of these stars.

\begin{figure}
\begin{center}
\includegraphics[angle=270,width=3.4in]{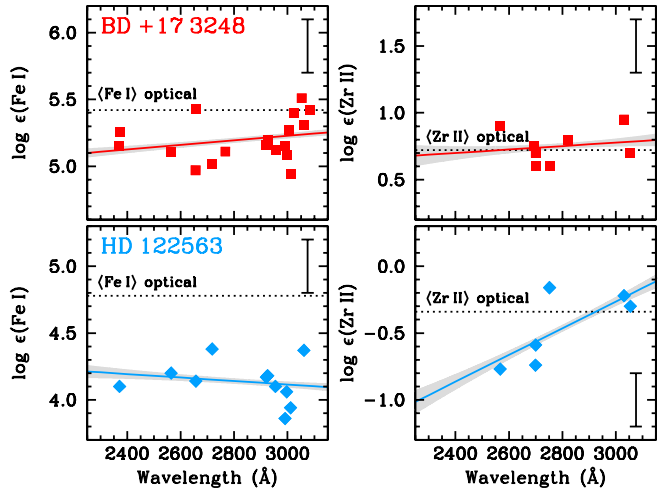}
\end{center}
\caption{
\label{fezr}
Abundances derived from NUV transitions of Fe~\textsc{i} and
Zr~\textsc{ii} in \bd\ and \hd.
A representative 1$\sigma$ abundance uncertainty for each
transition is illustrated.
The solid lines represent linear fits to the abundances,
with uncertainties indicated by the shaded regions.
The mean abundance of each species derived from transitions 
in the optical spectral range is indicated by the dotted lines.
Abundances have been renormalized to a common log($gf$) scale
in both stars, and the Zr~\textsc{ii} abundance in
\hd\ has been renormalized to the Zr~\textsc{ii} (optical)
abundance using the equivalent width measurements of \citet{honda04}
to account for the different model atmosphere parameters
between \citet{honda06} and the present study.
}
\end{figure}

Figure~\ref{fezr} displays the abundances of Fe~\textsc{i} and Zr~\textsc{ii}
in \bd\ and \hd\ as a function of wavelength.
Two characteristics would indicate that we have successfully
reproduced the continuous opacity:
(1) no trend between abundance and wavelength, and 
(2) agreement between the abundances derived from the 
optical and NUV transitions.
A similar phemonemon is observed in both \bd\ and \hd.
For Fe~\textsc{i}, we detect both an offset and a very slight trend
in both stars, though the effect is much smaller in 
\bd, the warmer of the two stars.
There is no offset and only a minimal trend for Zr~\textsc{ii} in \bd,
but a much larger trend is observed in \hd, although 
we have only derived abundances from 6 Zr~\textsc{ii} lines.
We use these trends as ``local metallicity'' references to
empirically adjust the derived abundances of other species
(where neutral species are adjusted according to Fe~\textsc{i}
and singly-ionized species are adjusted according to Zr~\textsc{ii}).
For example, the abundance derived from the Cd~\textsc{i} line at 2615\AA\
in \bd\ is adjusted by $+$0.32~dex, the difference between 
a hypothetical Fe~\textsc{i} line at 2615\AA\ and the mean Fe~\textsc{i}
abundance for lines in the optical spectral region.
We caution that there are very few Fe~\textsc{i} lines and no
Zr~\textsc{ii} lines shortward of the 
Mg~\textsc{i} series limit at 2515\AA, where the bound-free
opacity contribution from Mg~\textsc{i} may increase substantially.
This uncertainty should be borne in mind when extrapolating the trends to 
shorter wavelengths.

\begin{figure*}
\begin{center}
\includegraphics[angle=270,width=4.5in]{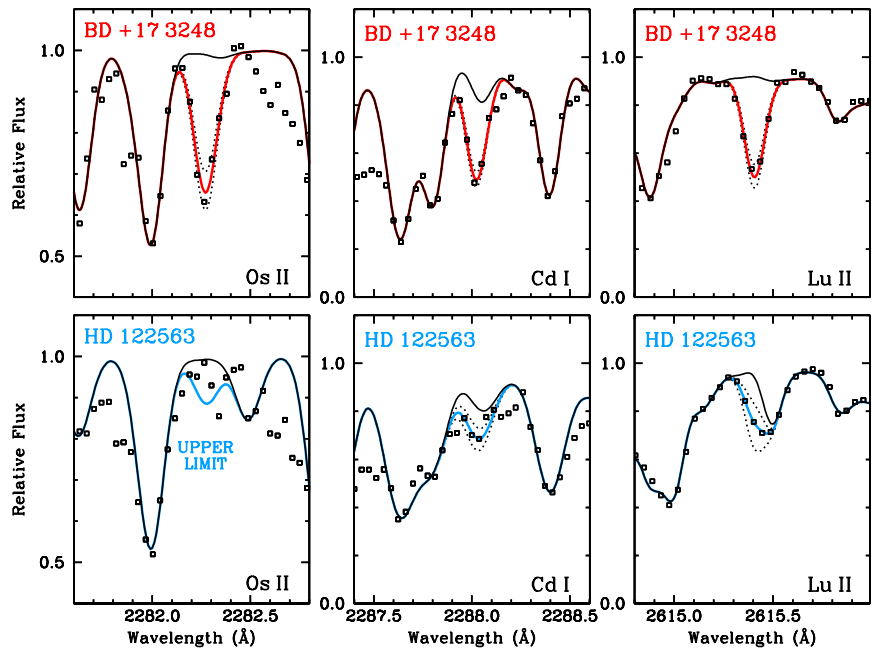}
\end{center}
\caption{
\label{specplot}
Fits of synthetic spectra to the observed spectra in \bd\ and \hd.
The bold lines indicate the best fit, dotted lines indicate
$\pm$~0.30~dex from the best fit, and the thin line indicates
a synthesis with the species of interest removed.
}
\end{figure*}

We derive the abundances of Cd~\textsc{i}, Lu~\textsc{ii}, and
Os~\textsc{ii} in \bd\ by comparing synthetic spectra to the
observed absorption profiles.
These fits are shown in Figure~\ref{specplot}, and the adjusted
abundances are reported in Table~\ref{abundtab}.
In \hd\ we report the tentative detection of 
Cd~\textsc{i} and Lu~\textsc{ii}, but 
we can only estimate an upper limit for Os~\textsc{ii}.
The Os~\textsc{ii} line is relatively unblended in \bd.
The Cd~\textsc{i} line is blended with 
an Fe~\textsc{i} transition at 2288.04\AA\ and an
As~\textsc{i} transition at 2288.12\AA.
Unfortunately, neither
has a laboratory log($gf$) measurement, so we are forced 
to fit these blends as best as possible.
The continuum is depressed slightly at the Lu~\textsc{ii} line in
\hd\ and more substantially in \bd\ by saturated Fe~\textsc{ii}
lines at 2613.82 and 2617.62\AA.
Furthermore, this line is contaminated with OH in \hd, although
this blend is minimized in the warmer atmosphere of \bd.
Considering all of these sources of uncertainty in the fits and
the corrections to account for the continuous opacity, we 
estimate an uncertainty of at least 0.30~dex for each
abundance derivation.

We have rederived the Hf~\textsc{ii} abundance in \bd\
using 4 transitions in the NUV.
Several Hf~\textsc{ii} transitions can also be detected in the 
optical spectral range. 
Finally, we have derived new or revised
abundances for several elements between the
1st and 2nd \rpro\ peaks in \bd\ (Nb, Mo, Ru, Rh, Pd, and Ag; 
$Z =$~41--42 and 44--47) using the Keck spectrum.
These abundances are reported in Table~\ref{abundtab}.

\section{Results and Discussion}
\label{results}

\begin{figure}
\begin{center}
\includegraphics[angle=0,width=3.4in]{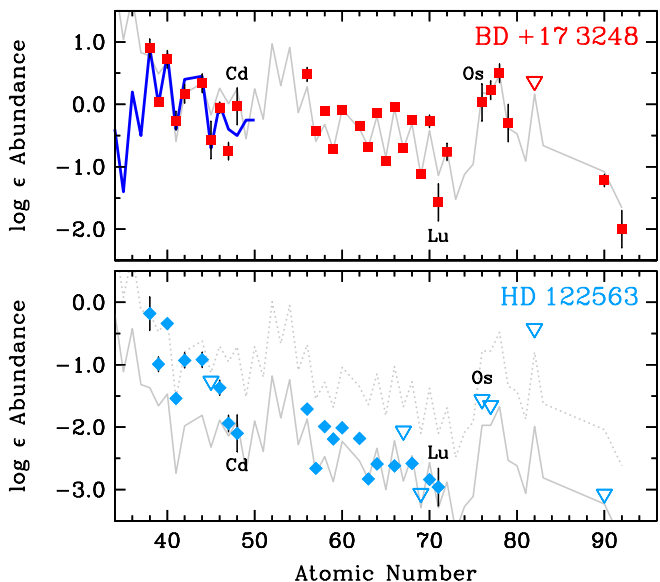}
\end{center}
\caption{
\label{solarr}
Neutron-capture abundance distributions in \bd\ and \hd.
Detections are indicated by filled symbols, and upper limits
are indicated by downward-facing open triangles.
The bold blue line in the top panel represents the core collapse
supernova high entropy neutrino wind calculations of 
\citet{farouqi09} (estimated from their Figure~3), normalized to 
Sr ($Z =$~38).
The solid line in each panel represents the scaled S.~S.\ \rpro\ abundance
distribution \citep{sneden08}, normalized to Eu ($Z =$~63),
and the dotted line in the lower panel represents this 
same distribution normalized instead to Sr.
Abundances are taken from \citet{cowan02,cowan05},
\citet{honda06}, \citet{roederer09}, \citet{sneden09},
and the present study.
Several elements have been renormalized to a common set of
laboratory log($gf$) values, and the abundances of
\hd\ have been renormalized to our abundance scale
as described in the caption of Figure~\ref{fezr}.
}
\end{figure}

In Figure~\ref{solarr} we display the abundance distribution for
the \ncap\ elements in \bd.
The S.~S.\ \rpro\ abundance distribution,
calculated as residuals from a classical model of the \spro\
\citep{sneden08}, is shown for comparison.
When normalized to the Eu abundance in \bd, this distribution
is a superb match to the stellar abundances for $Z \geq$~56, as has
been demonstrated previously (e.g., \citealt{cowan02}).

The detection of Cd extends the suite of lighter \ncap\ elements
observed in metal-poor stars more than halfway between the
1st and 2nd \rpro\ peaks (roughly $A \sim$~80 and 130, respectively).
For Sr to Cd ($Z =$~38--48, 
missing only the short-lived isotopes of Tc, $Z =$~43), there is a
very pronounced even-odd abundance pattern in \bd, much more so
than is predicted by the scaled S.~S.\ \rpro\ distribution
or the \rpro\ residuals derived from the
(Solar metallicity)
stellar \spro\ model of \citealt{arlandini99} (not shown).
This effect was first noticed in the Pd and Ag abundance pattern
of three \rpro\ enriched stars observed by \citet{johnson02}.
The large-scale dynamical network calculations from the
core collapse supernova high-entropy neutrino wind model of \citet{farouqi09}
reproduce this pattern better for Sr--Pd but still
underestimate the even-odd effect for Ag and Cd.
This general agreement is encouraging, but
additional calculations and comparisons are warranted.

Lu is the final member of the rare earth elements (REE) to be
unambiguously detected in \rpro\ enriched metal-poor stars.
The Lu/Eu (or, more generally, Lu/REE) 
ratio predicted by the scaled S.~S.\ \rpro\ distribution
is in reasonable, though not perfect, agreement with our
derived Lu abundance.
Additional Lu abundance derivations for other metal-poor, 
\rpro\ enhanced stars are required to assess 
whether the predicted Lu abundance or the stellar measurement
(or both) is in error.

Each of the neutral and singly-ionized states of Os have now been
detected in \bd, and our abundance of Os~\textsc{ii},
log~$\epsilon = +$0.03, 
is in fair agreement with an updated Os~\textsc{i} 
abundance derived from three optical and NUV lines, 
log~$\epsilon = +$0.25.
Os is the heaviest stable element that can be detected in its
singly-ionized state in \bd.
If the Os~\textsc{ii} abundance should prove reliable and its
uncertainty can be reduced, 
this has the potential to offer two significant improvements 
for nuclear cosmochronometry. 
The only radioactive isotopes practical for age dating the material
in old stars are \iso{232}{Th} and \iso{238}{U}, both of which
can only be detected as first ions.
Abundance uncertainties are minimized when considering 
ratios of two elements in the same ionization state.
When predicting the initial production ratios, 
the uncertainty is generally smallest when the two elements
are as close in mass number as possible.
Previously, the heaviest singly-ionized reference 
element has been Hf~\textsc{ii}, whose stable isotopes are
separated by 52--55 mass units from \iso{232}{Th};
the stable \rpro\ isotopes of Os are only separated by 40--44 mass units
from \iso{232}{Th}.
Adopting the range of production ratios from \citet{kratz07},
the Th~\textsc{ii}/Os~\textsc{i} ratio predicts an age range of 
15.7--21.5~Gyr,
whereas Th~\textsc{ii}/Os~\textsc{ii} predicts an age range of 
5.4--11.2~Gyr.
The latter is in better agreement with the age predicted from other
chronometer pairs 
(e.g., Th~\textsc{ii}/Eu~\textsc{ii}, which predicts an age range of 
7.9--12.3~Gyr).
The present uncertainty in our Os~\textsc{ii} abundance translates to an
age precision of 14~Gyr, but if the Os~\textsc{ii} uncertainty 
could be reduced to 0.10~dex 
the age precision would improve to 4.7~Gyr.

The Hf~\textsc{ii} abundance derived in \bd\ from 4 lines in the 
NUV is marginally lower 
(log~$\epsilon = -$0.76~$\pm$~0.08, $\sigma =$~0.14)
than that derived from 6 lines in the optical
(log~$\epsilon = -$0.57~$\pm$~0.03, $\sigma =$~0.08).
Previous analyses of the Hf~\textsc{ii} abundance in 
\rpro\ enriched metal-poor stars have revealed that the 
stellar Hf~\textsc{ii} \rpro\ abundance is higher
by 0.15--0.25~dex
than that predicted by the scaled S.~S.\ \rpro\ distribution
\citep{lawler07,roederer09,sneden09}.
Our Hf~\textsc{ii} NUV abundance is in excellent agreement with
the scaled S.~S.\ \rpro\ Hf/REE ratio.
Many of the transitions used to derive the REE stellar \rpro\ 
abundance distribution from the optical spectral range 
arise from 0.0~eV lower levels, but only one of the 12 Hf~\textsc{ii}
lines employed by \citet{lawler07} has a 0.0~eV lower level.
Two of the four lines used to derive our 
NUV Hf~\textsc{ii} abundance arise from 0.0~eV levels,
with log~$\epsilon = -$0.68 
from just these two transitions.
Perhaps by using these transitions we have 
mitigated a subtle systematic effect present in the computation 
of the Hf~\textsc{ii} abundance relative to other REE.
This might imply that other stellar Hf~\textsc{ii} \rpro\ 
abundances---rather than the predicted S.~S.\ \rpro\
abundances---warrant minor revisions.
Again adopting the range of production ratios from \citet{kratz07},
the NUV Th~\textsc{ii}/Hf~\textsc{ii} ratio predicts an age range of 
5.8--18.5~Gyr, whereas the optical 
Th~\textsc{ii}/Hf~\textsc{ii} ratio predicts an age of 
14.7--27.4~Gyr.
The precision is 5--6~Gyr in each measurement, but clearly
the lower Hf~\textsc{ii} abundance derived from the NUV 
lines provides a more realistic age estimate for \bd.

Figure~\ref{solarr} also displays the abundance distribution
for \hd, which is known to be deficient in the heavy \ncap\ elements.
The scaled S.~S.\ \rpro\
distribution is a poor fit to the abundance pattern
whether normalized to the 1st \rpro\ peak or the REE
\citep{honda06}.
(No reasonable \spro\ distribution, or combination of $r$- and
\spro\ distributions, matches either.)
\hd\ may be an example of
enrichment by the so-called ``weak'' \rpro, which produces small amounts
of light \ncap\ material and steadily-decreasing amounts of heavier material
\citep{honda06,wanajo06}.
Our Cd abundance in \hd\ suggests that the downward abundance trend 
continues in the region between the 1st and 2nd \rpro\ peaks.
Our Os upper limit in this star is not strong enough
to exclude a scaled S.~S.\ \rpro\ pattern between the REE
and the 3rd \rpro\ peak. 

The detection of these three new species in \bd\ is only
a first step in understanding how and in what amount 
these elements were produced.
By examining their abundances in other
metal-poor stars enriched to different levels by the \rpro, we may
gain a better sense of any systematic offsets 
affecting the present analysis.
These uncertainties must be minimized to take full advantage
of these species as constraints on \ncap\ nucleosynthesis models
and meaningful age probes for the \ncap\ material in metal-poor stars.

\acknowledgments

We thank the referee, David Lai, for a careful review of the manuscript.
The authors wish to recognize and acknowledge the very significant 
cultural role and reverence that the summit of Mauna Kea has always 
had within the indigenous Hawaiian community.  We are most fortunate 
to have the opportunity to conduct observations from this mountain.
This research has made use of the 
NASA Astrophysics Data System (ADS) and the
NIST Atomic Spectra Database.
Funding for this project has been generously provided by 
the U.~S.\ National Science Foundation
(grants AST~09-08978 to C.S., 
AST~09-07732 to J.E.L., and 
AST~07-07447 to J.J.C.).


\begin{deluxetable}{cccccccc}
\tablecaption{Derived Stellar Abundances
\label{abundtab}}
\tablewidth{0pt}
\tablehead{
\colhead{} &
\colhead{} &
\colhead{} &
\colhead{} &
\colhead{} &
\colhead{} &
\colhead{BD~$+$17~3248} &
\colhead{HD~122563} \\
\colhead{Species} &
\colhead{Z} &
\colhead{$\lambda$ (\AA)} &
\colhead{E.P.\ (eV)} &
\colhead{log($gf$)} &
\colhead{Ref.} &
\colhead{log~$\epsilon$} &
\colhead{log~$\epsilon$}}
\startdata
\hline
\multicolumn{8}{c}{New Abundances from STIS Spectra} \\
\hline
Cd~\textsc{i}                   & 48 & 2288.02 & 0.00 & $+$0.15 & 1 & $-$0.03 & $-$2.10   \\
Lu~\textsc{ii}                  & 71 & 2615.41 & 0.00 & $+$0.11 & 2 & $-$1.58 & $-$2.96   \\
Hf~\textsc{ii}\tablenotemark{a} & 72 & 2638.72 & 0.00 & $-$0.17 & 3 & $-$0.62 & \nodata   \\
Hf~\textsc{ii}\tablenotemark{a} & 72 & 2641.41 & 1.04 & $+$0.57 & 3 & $-$0.91 & \nodata   \\
Hf~\textsc{ii}\tablenotemark{a} & 72 & 2820.23 & 0.38 & $-$0.05 & 3 & $-$0.77 & \nodata   \\
Hf~\textsc{ii}\tablenotemark{a} & 72 & 3012.90 & 0.00 & $-$0.60 & 3 & $-$0.74 & \nodata   \\
Os~\textsc{ii}                  & 76 & 2282.28 & 0.00 & $-$0.14 & 4 & $+$0.03 & $< -$1.56 \\
\hline
\multicolumn{8}{c}{New Abundances from HIRES Spectra} \\
\hline
Nb~\textsc{ii}                  & 41 & 3215.59 & 0.44 & $-$0.24 & 5 & $-$0.26 & \nodata   \\
Mo~\textsc{i}                   & 42 & 3864.10 & 0.00 & $-$0.01 & 6 & $+$0.17 & \nodata   \\
Ru~\textsc{i}                   & 44 & 3498.94 & 0.00 & $+$0.31 & 7 & $+$0.34 & \nodata   \\
Rh~\textsc{i}                   & 45 & 3434.89 & 0.00 & $+$0.45 & 8 & $-$0.57 & \nodata   \\
Pd~\textsc{i}\tablenotemark{a}  & 46 & 3242.70 & 0.81 & $+$0.07 & 9 & $-$0.10 & \nodata   \\
Pd~\textsc{i}\tablenotemark{a}  & 46 & 3404.58 & 0.81 & $+$0.33 & 9 & $-$0.09 & \nodata   \\
Pd~\textsc{i}\tablenotemark{a}  & 46 & 3516.94 & 0.94 & $-$0.21 & 9 & $+$0.05 & \nodata   \\
Ag~\textsc{i}\tablenotemark{a}  & 47 & 3280.67 & 0.00 & $-$0.04 &10 & $-$0.83 & \nodata   \\
Ag~\textsc{i}\tablenotemark{a}  & 47 & 3382.90 & 0.00 & $-$0.35 &10 & $-$0.67 & \nodata   \\
\enddata
\tablenotetext{a}{
The mean abundances in \bd\ are 
\eps{Ag~\textsc{i}}~$= -0.75 \pm 0.14$,
\eps{Pd~\textsc{i}}~$= -0.05 \pm 0.09$, and
\eps{Hf~\textsc{ii}}~$= -0.76 \pm 0.14$.
}
\tablerefs{
(1)  \citet{morton00};
(2)  this study;
(3)  \citet{lawler07};
(4)  \citet{ivarsson04};
(5)  \citet{nilsson10};
(6)  \citet{whaling88};
(7)  \citet{wickliffe94};
(8)  \citet{kwiatkowski82};
(9)  \citet{xu06};
(10) \citet{fuhr09}
}
\end{deluxetable}

\end{document}